\documentclass[twocolumn,pra,showpacs,showkeys,amsmath,amssymb]{revtex4}
\usepackage{amsthm,graphics,graphicx}

\theoremstyle{plain}
\newtheorem{Lemm}{Lemma}
\newtheorem{Thrm}{Theorem}

\begin{document}
\title{Quantum String Seal Is Insecure}
\author{H.~F. Chau}
\email{hfchau@hkusua.hku.hk}
\affiliation{Department of Physics, University of Hong Kong, Pokfulam Road,
 Hong Kong}
\affiliation{Center of Theoretical and Computational Physics, University of
 Hong Kong, Pokfulam Road, Hong Kong}
\date{\today}

\begin{abstract}
 A quantum string seal encodes the value of a (bit) string as a quantum state
 in such a way that everyone can extract a non-negligible amount of available
 information on the string by a suitable measurement. Moreover, such
 measurement must disturb the quantum state and is likely to be detected by an
 authorized verifier. In this way, the intactness of the encoded quantum state
 plays the role of a wax seal in the digital world. Here I analyze the security
 of quantum string seal by studying the information disturbance tradeoff of a
 measurement. This information disturbance tradeoff analysis extends the
 earlier results of Bechmann-Pasquinucci \emph{et al.} and Chau by concluding
 that all quantum string seals are insecure. Specifically, I find a way to
 obtain non-trivial available information on the string that escapes the
 verifier's detection with at least 50\% chance.
\end{abstract}

\maketitle
\section{Introduction}
\label{Sec:Intro}
 The idea of quantum seal was introduced by Bechmann-Pasquinucci to capture the
 essence of an envelop with sealed wax in the digital world \cite{quantseal}.
 Specifically, a quantum (bit) seal is a method to encode a classical bit by
 some quantum particles in such a way that everyone can find out the value of
 the bit with high chance by an appropriate measurement. Moreover, any such
 measurement must disturb the state of the quantum particles so that an
 authorized person, who has some extra information on the state of the
 particles, can detect such a measurement with high probability
 \cite{quantseal}.

 A few quantum sealing schemes have been proposed. They fall into the following
 three types. The original scheme by Bechmann-Pasquinucci \cite{quantseal} and
 the one by Chau \cite{oldseal} are perfect quantum bit seals because everyone
 can find out the value of the bit with certainty. The quantum bit seals
 introduced by Singh and Srikanth \cite{anotherseal} as well as He
 \cite{anotherseal2} are imperfect quantum bit seals as readers cannot
 correctly determine the value of the bit with certainty. Recently, He
 \cite{stringseal} extended the notion of sealing a bit to sealing a bit string
 in a natural way by proposing an imperfect quantum bit string seal which
 separately encodes each bit of a classical string. Parameters in He's scheme
 are set in such a way that one has a high chance to correctly extract a large
 portion of the string. However, one has negligibly small chance to correctly
 determine the whole string \cite{stringseal}.

 Density matrices representing any two distinct classical messages in a perfect
 quantum seal must be orthogonal. Hence, there exits a projective measurement
 to find out the encoded message without disturbing the quantum state. Using
 this idea, Bechmann-Pasquinucci \emph{et al.} proved the insecurity of all
 perfect quantum (bit) seals provided that one has access to a quantum computer
 \cite{impossibleperfect}. Recently, He showed certain bounds relating the
 information gain and the measurement detection probability for any quantum bit
 seal \cite{bound}. However, his bound is not tight. In contrast, Chau proved
 that all imperfect quantum bit seals are insecure by giving an explicit
 measurement strategy. In addition, he obtained a lower bound for the fidelity
 of the resultant quantum state after a measurement. More importantly, this
 lower bound is greater than 1/2 and is attainable by certain quantum sealing
 schemes \cite{bitsealimposs}.

 The insecurity proof by Chau in Ref.~\cite{bitsealimposs} relies heavily on
 the properties of trace distance between two density matrices. Generalizing
 his proof to the case of sealing more than two classical states is not
 straight-forward. Furthermore, the security analysis of quantum string
 commitment by Buhrman \emph{et al.}
 \cite{bitstringcommitment,bitstringcommitment2} illustrated two important
 points. First, more than one inequivalent security parameters may exist for a
 quantum string cryptographic scheme; and second, the security of a quantum
 string cryptographic scheme may be very different from that of a quantum bit
 one. Therefore, it is important to study the security of quantum string seal
 thoroughly.

 In this Paper, I analyze the information disturbance tradeoff for a general
 quantum seal that maps a fixed number $N$ of distinct classical messages to
 $N$ density matrices. Moreover, the probabilities of occurrence of these
 classical messages need not be equal. I begin by introducing the general
 formalism and the security requirements for a quantum seal as well as the
 notion of the most stringent quantum seal in Sec.~\ref{Sec:QSeal}. Then I
 report a measurement strategy that obtains non-zero amount of information on
 the original message and introduce two performance measures in
 Sec.~\ref{Sec:Optimal}. I also prove the optimality of this measurement
 strategy against the most stringent quantum seal under one of the performance
 measures. And I also argue that this strategy also performs well under the
 other performance measure in Sec.~\ref{Sec:Optimal}. Using these information
 disturbance analyses, I show that all imperfect quantum seals, including
 quantum string seals in which $N = 2^n$, are insecure in
 Sec.~\ref{Sec:Insecurity}. Finally, a summary is given in
 Sec.~\ref{Sec:Conclusion}.
 
\section{Quantum Seal}
\label{Sec:QSeal}
\subsection{Quantum seal and its security requirements}
\label{Subsec:SecurityRequirement}
 A quantum seal is a scheme for Alice to encode a fixed number $N$ of distinct
 classical messages as publicly accessible quantum mixed states known as the
 sealed states. (Although $N$ is a fixed integer and the dimension of the
 sealed states is finite, readers can check that the proofs and discussions
 reported in this paper can be easily extended to the case when the number of
 distinct classical messages $N$ as well as the dimension of the quantum mixed
 states involved are infinite.) It has to satisfy:
\begin{enumerate}
 \item Any member of the public, say Bob, can correctly determine a significant
  portion of the original classical message chosen by Alice, known as the
  sealed message, with non-negligible probability by a measurement.
  \label{Item:mutual_inform}
 \item Any such measurement in criterion~\ref{Item:mutual_inform} must disturb
  the sealed state so that an authorized verifier may correctly detect the
  measurement with non-zero (unconditional) probability.
  \label{Item:detect_prob}
 \item Criterion~\ref{Item:detect_prob} still holds after replacing the
  unconditional probability by the probability conditioned on Bob's successful
  determination of a significant portion of the sealed message.
  \label{Item:detect_prob_cond}
\end{enumerate}

 In the language of information theory, criterion~\ref{Item:mutual_inform}
 means that the mutual information on the sealed message obtained by Bob 
 divided by the entropy of the original classical message is of order of $1$.
 Clearly, this criterion is necessary; for otherwise, even an honest Bob has
 negligible chance to obtain a significant fraction of the sealed message.

 The maximum probability for Bob to correctly determine the sealed message can
 be made close to $1/N$. Nonetheless, in order to satisfy
 criterion~\ref{Item:mutual_inform}, there exists a partition ${\mathfrak P}$
 of the $N$ distinct classical messages with $\log |{\mathfrak P}| \lesssim
 \log N$ such that the maximum probability of correctly determining which set
 in the partition the original message belongs to is much greater than
 $1/|{\mathfrak P}|$. (Let's use He's quantum string scheme as an example to
 illustrate this point. Although it is not likely to correctly extract the
 entire sealed string in He's scheme, it is highly probable to correctly
 determine, say, the first 99\% of the sealed string \cite{stringseal}. In
 fact, one possible choice of ${\mathfrak P}$ in this case is to partition
 according to the values of the first 99\% of the original message string.)
 Through the partition ${\mathfrak P}$, the quantum seal that encodes $N$
 distinct classical messages can be regarded as a seal that encodes
 $|{\mathfrak P}|$ distinct classical messages. With this identification in
 mind, I may assume from now on that the maximum probability of correctly
 finding the original classical message $p_\text{max}$ to be much greater than
 $1/N$. Furthermore, without loss of generality, I label these $N$ classical
 messages by $0,1,\ldots ,N-1$ in such a way that the \emph{a priori}
 probability of occurrence $\eta_i$ for the message $i$ obeys the constraints
 $\eta_0 \geq \eta_1 \geq \cdots \geq \eta_{N-1} > 0$ and $\sum_{i=0}^{N-1}
 \eta_i = 1$.
 
 A quantum sealing scheme is called a \emph{quantum bit seal} if $N=2$ and a
 \emph{quantum (bit) string seal} if $N = 2^n$. Moreover, the scheme is
 \emph{perfect} if Bob can determine the entire classical message with
 certainty; otherwise, the scheme is \emph{imperfect} \cite{bitsealimposs}.

\subsection{A game-theoretic formulation of the problem}
\label{Subsec:Game_Formation}
 Quantum seal is in some sense a game between Alice and Bob. For a given
 sealing scheme used by Alice, Bob tries to gain as much information on the
 sealed message as possible on the one hand and to reduce the chance of being
 caught by the verifier on the other hand. And Alice surely wants to pick a
 sealing scheme that makes Bob's task as difficult as possible.

 To analyze the security of quantum seals, all sealed states can be assumed to
 be pure as using purified states increase the verifier's chance to detect
 Bob's measurement \cite{bitsealimposs}. Using the notation used in
 Ref.~\cite{bitsealimposs}, the sealed state for message $i$ is
\begin{equation}
 |\tilde{\psi}_i\rangle = \sum_j \lambda_{ij} |\psi_{ij}\rangle_\text{B}
 \otimes |\phi_j\rangle_\text{A} \label{E:Def_psi_ij}
\end{equation}
 for all $i$, where $|\psi_{ij}\rangle$'s are normalized states that are not
 necessarily mutually orthogonal, and $|\phi_j\rangle$'s are orthonormal
 states. Note that particles labeled by the subscript ``B'' in
 Eq.~(\ref{E:Def_psi_ij}) are publicly accessible, and those labeled by the
 subscript ``A'' are accessible only to authorized verifiers. Thus, the state
 of the publicly accessible particles is
\begin{equation}
 \rho_i = \sum_j |\lambda_{ij} |^2 |\psi_{ij}\rangle\langle\psi_{ij}| ~.
 \label{E:Def_rho_i}
\end{equation}
 if the sealed message is $i$.

 Bob's attempt to obtain some information on the classical message can be
 described by a positive operator-valued measure (POVM) measurement
 ${\mathcal E}$ on the publicly accessible particles. From the verifier's point
 of view, this measurement changes the sealed state $|\tilde{\psi}_i\rangle
 \langle\tilde{\psi}_i|$ to ${\mathcal E} \otimes I (|\tilde{\psi}_i\rangle
 \langle\tilde{\psi}_i|) \equiv \tilde{\mathcal E} (|\tilde{\psi}_i\rangle
 \langle\tilde{\psi}_i|)$.

 I write ${\mathcal E} = \sum_{j=0}^{N-1} {\mathcal L}_j$, where
 ${\mathcal L}_j$ is the superoperator describing Bob's action when he
 concludes that the sealed message is $j$. Then, the probability that Alice's
 original message is $i$ and Bob's measurement on the sealed state yields $j$
 is given by
\begin{equation}
 \text{Pr}_{ij} = \eta_i \text{Tr} [ {\mathcal L}_j \otimes I (|\tilde{\psi}_i
 \rangle\langle\tilde{\psi}_i |) ] \equiv \eta_i \text{Tr} [
 \tilde{\mathcal L}_j (|\tilde{\psi}_i \rangle\langle\tilde{\psi}_i |) ] ~.
 \label{E:Def_Pr_ij}
\end{equation}
 In addition, Bob's mutual information on the original message equals
\begin{eqnarray}
 {\mathcal I} & \equiv & {\mathcal I}({\mathcal E}) \equiv {\mathcal I}
  (\tilde{\mathcal E}) \nonumber \\
 & = & -\sum_i \eta_i \log_2 \eta_i - \sum_j \left( \sum_k \text{Pr}_{kj}
  \right) \log_2 \left( \sum_k \text{Pr}_{kj} \right) \nonumber \\
 & & ~~+ \sum_{i,j} \text{Pr}_{ij} \log_2 \text{Pr}_{ij} ~.
 \label{E:mutual_inform}
\end{eqnarray}

 A simple way to measure Bob's average chance of being caught is to compute the
 average fidelity of the sealed state \cite{bitsealimposs}, namely,
\begin{equation}
 \bar{F} \equiv \bar{F}({\mathcal E}) \equiv \bar{F}(\tilde{\mathcal E}) =
 \sum_{i=0}^{N-1} \eta_i \langle \tilde{\psi}_i | \tilde{\mathcal E}
 (|\tilde{\psi}_i\rangle\langle \tilde{\psi}_i|) | \tilde{\psi}_i \rangle ~.
 \label{E:Def_F}
\end{equation}
 In other words, the average fidelity (or fidelity for short) $\bar{F}$
 measures the mean disturbance caused by Bob's measurement ${\mathcal E}$.
 Since $1 - \bar{F}$ is the probability of detecting Bob's measurement,
 $\bar{F}$ is a performance indicator for criterion~\ref{Item:detect_prob}
 stated in Subsec.~\ref{Subsec:SecurityRequirement}.

 Bob's chance of being caught given that he correctly determined the sealed
 message is reflected in the average fidelity of the sealed state conditional
 on Bob's success $\bar{F}_\text{cond}$, namely,
\begin{eqnarray}
 \bar{F}_\text{cond} & \equiv & \bar{F}_\text{cond}({\mathcal E}) \equiv
  \bar{F}_\text{cond}(\tilde{\mathcal E}) \nonumber \\
 & = & \sum_{i=0}^{N-1} \frac{\eta_i \langle\tilde{\psi}_i |
  \tilde{\mathcal L}_i (|\tilde{\psi}_i\rangle\langle\tilde{\psi}_i |) |
  \tilde{\psi}_i \rangle}{\text{Tr}[ \tilde{\mathcal L}_i (|\tilde{\psi}_i
  \rangle\langle\tilde{\psi}_i |)]} ~. \label{E:Def_F_cond}
\end{eqnarray}
 (Note that those terms with $\text{Tr} [ \tilde{\mathcal L}_i(|\tilde{\psi}_i
 \rangle\langle\tilde{\psi}_i |)] = 0$ in the above equation are regarded as
 $0$.) Thus, the average conditional fidelity (or conditional fidelity for
 short) $\bar{F}_\text{cond}$ measures the mean disturbance caused by Bob's
 measurement ${\mathcal E}$ conditioned on his successful recovery of the
 sealed message. Since $1 - \bar{F}_\text{cond}$ is the probability of
 detecting Bob's measurement conditioned on Bob's successful recovery of the
 sealed message, $\bar{F}_\text{cond}$ is a performance indicator for
 criterion~\ref{Item:detect_prob_cond} in
 Subsec.~\ref{Subsec:SecurityRequirement}.

 I denote the probability of correctly determining the sealed message using the
 POVM measurement ${\mathcal E}$ by $p$. Surely, $p \leq p_\text{max}$.
 Furthermore, I assume $p\geq 1/N$ as no one is interested in those
 ${\mathcal E}'s$ that perform worse than random guessing. (Interestingly, the
 scheme with $\eta_i = 1/N$ and $\rho_i = \rho_j$ for all $i,j$ illustrates
 that $p$ may not be less than $1/N$.)

 Amongst all the POVM's whose probability of correctly determining the sealed
 message is $p$, Bob would like to pick the one that minimizes the average
 chance of being detected by a verifier (and hence maximizes $\bar{F}$). In
 contrast, Alice would like to pick a seal that minimizes $\bar{F}$. Thus, the
 average fidelity $\min_\text{Alice} \max_\text{Bob} \bar{F}(p,p_\text{max})$
 for the optimal measurement strategy against the most stringent seal is found
 by first taking the maximum over all possible POVM measurements
 ${\mathcal E}$'s used by Bob whose probability of correctly determining the
 original message is $p$ for a given sealing scheme, and then by taking the
 minimum over all quantum seals with maximum probability of correctly
 determining the message $p_\text{max}$ by Alice \cite{bitsealimposs}. The
 average conditional fidelity for the optimal measurement strategy against the
 most stringent seal $\min_\text{Alice} \max_\text{Bob} \bar{F}_\text{cond}(p,
 p_\text{max})$ is similarly defined.

 Since $\bar{F}$ and $\bar{F}_\text{cond}$ are two different performance
 indicators, one expects that $\min_\text{Alice} \max_\text{Bob} \bar{F}$ and
 $\min_\text{Alice} \max_\text{Bob} \bar{F}_\text{cond}$ have to be attained by
 two different sealing schemes and measurement strategies. I show in
 Sec.~\ref{Sec:Optimal} that this is indeed the case for a general $N$.
 Nevertheless, for $N \leq 5$, optimal sealing scheme and measurement
 strategies as reflected by the two performance indicators can be chosen to be
 the same.

\section{The Optimal Measurement Strategy}
\label{Sec:Optimal}
\subsection{A measurement strategy and its performance}
\label{Subsec:POVM}
 Obviously, $p_\text{max}$ is equal to the maximum probability of correctly
 distinguishing the mixed states $\rho_i$'s with \emph{a priori} occurrence
 probabilities $\eta_i$'s. The set of POVM elements $\{ \Pi_k \}_{k=0}^{N-1}$
 that maximizes such a probability is given by \cite{opt1,opt2}
\begin{equation}
 \Pi_k (\eta_k \rho_k - \eta_j \rho_j) \Pi_j = 0 \label{E:opt_dist1}
\end{equation}
 and
\begin{equation}
 \sum_i \eta_i \rho_i \Pi_i - \eta_j \rho_j \geq 0 \label{E:opt_dist2}
\end{equation}
 for all $0\leq j,k\leq N-1$. In other words, $p_\text{max}$ and $\Pi_i$'s
 satisfy the equation
\begin{equation}
 p_\text{max} = \sum_{i=0}^{N-1} \eta_i \text{Tr} (\Pi_i \rho_i) =
 \sum_{i=0}^{N-1} \eta_i \langle\tilde{\psi}_i | \Pi_i \otimes I |
 \tilde{\psi}_i \rangle ~. \label{E:Def_p_max}
\end{equation}
 Clearly, $1/N \leq \eta_0 \leq p_\text{max} \leq 1$. In fact, $p_\text{max} =
 1/N$ if and only if $\rho_i = \rho_j$ for all $i,j$; and $p_\text{max} = 1$ if
 and only if $\rho_i$'s are mutually orthogonal.

 I write the spectral decomposition of $\Pi_i$ as
\begin{equation}
 \Pi_i = \sum_j \mu_{ij} |e_{ij}\rangle\langle e_{ij} | ~, \label{E:Pi_i_Spec}
\end{equation}
 where $\{ |e_{ij}\rangle \}_j$ are complete sets of orthonormal state kets for
 all $i$ and $\mu_{ij} \geq 0$ for all $i,j$. Based on the $\Pi_i$'s, I
 construct
\begin{equation}
 M_i = \sum_j \left( \frac{1-\nu}{N} + \nu \mu_{ij} \right)^{1/2} |e_{ij}
 \rangle\langle e_{ij}| ~, \label{E:Def_M_i}
\end{equation}
 where
\begin{equation}
 \nu \equiv \nu (p,p_\text{max},N) = \frac{p N - 1}{p_\text{max} N - 1}
 \label{E:Def_nu}
\end{equation}
 for all $p\in [1/N,1]$. Clearly, $\nu\in [0,1]$ and hence $M_i$'s are
 well-defined measurement operators. I denote the POVM measurement with Kraus
 operators $\{ M_i \}_{i=0}^{N-1}$ by ${\mathcal E}_{p,p_\text{max}}$ for
\begin{equation}
 \sum_{i=0}^{N-1} \eta_i \text{Tr} (M_i^\dag M_i \rho_i) = \frac{1-\nu}{N} +
 \nu \sum_{i=0}^{N-1} \eta_i \text{Tr} (\Pi_i \rho_i) = p ~. \label{E:M_i_p}
\end{equation}

 Using the POVM measurement ${\mathcal E}_{p,p_\text{max}}$, the probability
 that Alice's original classical message is $i$ and Bob's measurement on the
 sealed state yields $j$ is equal to
\begin{eqnarray}
 \text{Pr}_{ij} & = & \eta_i \langle\tilde{\psi}_i | ( M^\dag_j \otimes I ) (
  M_j \otimes I ) | \tilde{\psi}_i \rangle \nonumber \\
 & = & \eta_i \left[ \frac{1-\nu}{N} + \nu \text{Tr} (\Pi_j \rho_i) \right] ~.
 \label{E:Pr_ij} 
\end{eqnarray}
 In particular, if $p, p_\text{max}$ are large and $\nu$ is close to $1$, then
 $\text{Pr}_{ii}$ is generally much larger than $\text{Pr}_{ij}$ for $j\neq i$.
 Consequently, the mutual information ${\mathcal I}$ is close to the maximum
 possible value of $-\sum_i \eta_i \log_2 \eta_i$. (For example, in the He's
 scheme \cite{stringseal}, Bob's mutual information obtained by the POVM
 ${\mathcal E}_{p_\text{max},p_\text{max}}$ on the sealed message equals $I =
 0.99n [1 + \epsilon \log_2 \epsilon + (1-\epsilon) \log_2 \epsilon]$ where
 $\epsilon$ is the small control parameter in his scheme.)

 To investigate the disturbance caused by this POVM measurement, I use the
 following lemma.

\begin{Lemm}
 Let $0 \leq \nu \leq 1$. Then,
 \begin{eqnarray}
  f(x) & = & \sqrt{\nu x + \frac{1-\nu}{N}} \,-\, \sqrt{\frac{1-\nu}{N}}
   \nonumber \\
  & & - \left( \sqrt{\nu + \frac{1-\nu}{N}} - \sqrt{\frac{1-\nu}{N}} \right) x
   \geq 0 \label{E:lemma_inequality}
 \end{eqnarray}
 for all $x \in [0,1]$. Besides, the equality holds if and only if $x = 0$ or
 $1$. \label{Lemm:inequality}
\end{Lemm}
\noindent \emph{Proof.}
 By solving the equation $df/dx = 0$ and considering $d^2 f/dx^2$, I find that
 the continuous function $f(x)$ has a single local maximum in the interval
 $[0,1]$. Hence, $f(x) \geq \min ( f(0),f(1) ) = 0$ for all $x\in [0,1]$.
 Moreover, $f(x) = 0$ if and only if $x = 0$ or $1$.
\hfill$\Box$

\par\medskip
 A direct consequence of Lemma~\ref{Lemm:inequality} is that
\begin{eqnarray}
 M_i & \geq & \sqrt{\frac{1-\nu}{N}} I + \left( \sqrt{\nu + \frac{1-\nu}{N}} -
  \sqrt{\frac{1-\nu}{N}} \right) \Pi_i \nonumber \\
 & \equiv & a(\nu,N) I + b(\nu,N) \Pi_i \equiv a I + b \Pi_i
 \label{E:operator_inequality}
\end{eqnarray}
 for all $i$. And the equality holds if and only if $\Pi_i$ is a projector.

 To find a lower bound for $\min_\text{Alice} \max_\text{Bob} \bar{F}$, I
 substitute Eq.~(\ref{E:operator_inequality}) into Eq.~(\ref{E:Def_F}) to
 obtain
\begin{eqnarray}
 & & \min_\text{Alice} \max_\text{Bob} \bar{F} (p,p_\text{max}) \nonumber \\
 & \geq & \bar{F}({\mathcal E}_{p,p_\text{max}}) \nonumber \\
 & \geq & \sum_{i,j=0}^{N-1} \eta_i \left| \langle \tilde{\psi}_i | a I \otimes
  I + b \Pi_j \otimes I | \tilde{\psi}_i \rangle \right|^2 \nonumber \\
 & = & N a^2 + 2 a b + b^2 \sum_{i,j=0}^{N-1} \eta_i \left| \langle
  \tilde{\psi}_i | \Pi_j \otimes I | \tilde{\psi}_i \rangle \right|^2 ~.
 \label{E:Fbar_inequality}
\end{eqnarray}
 Subjected to the constraints in Eq.~(\ref{E:Def_p_max}) and
\begin{equation}
 \sum_{j=0}^{N-1} \langle \tilde{\psi}_i | \Pi_j \otimes I | \tilde{\psi}_i
 \rangle = 1 \label{E:OperatorSumConstraint}
\end{equation}
 for all $i$, the last line of Eq.~(\ref{E:Fbar_inequality}) is minimized if
\begin{equation}
 \langle \tilde{\psi}_i | \Pi_j \otimes I | \tilde{\psi}_i \rangle = \left\{
 \begin{array}{cl}
  p_\text{max} & \text{if~} i = j ~, \\
  ~ \\
  \displaystyle \frac{1-p_\text{max}}{N-1} & \text{if~} i \neq j ~.
 \end{array}
 \right. \label{E:minimize_condition}
\end{equation}
 Consequently,
\begin{eqnarray}
 & & \min_\text{Alice} \max_\text{Bob} \bar{F}(p,p_\text{max}) \nonumber \\
 & \geq & 1 - \left( \sqrt{\nu+\frac{1-\nu}{N}} - \sqrt{\frac{1-\nu}{N}}
  \right)^2 \times \nonumber \\
 & & ~~~\left[ 1 - p_\text{max}^2 - \frac{(1-p_\text{max})^2}{N-1} \right]
  \label{E:opt_F}
\end{eqnarray}
 for all $1/N \leq p \leq p_\text{max}$, where $\nu$ is given by
 Eq.~(\ref{E:Def_nu}).
 
 To find a lower bound for $\min_\text{Alice} \max_\text{Bob}
 \bar{F}_\text{cond}$, I substitute Eq.~(\ref{E:operator_inequality}) into
 Eq.~(\ref{E:Def_F_cond}) to obtain
\begin{eqnarray}
 & & \min_\text{Alice} \max_\text{Bob} \bar{F}_\text{cond}(p,p_\text{max})
  \nonumber \\
 & \geq & \bar{F}_\text{cond} ({\mathcal E}_{p,p_\text{max}}) \nonumber \\
 & \geq & \sum_{i=0}^{N-1} \frac{\eta_i \left( a + b \langle \tilde{\psi}_i |
  \Pi_i \otimes I | \tilde{\psi}_i \rangle \right)^2}{a^2 + \nu \langle
  \tilde{\psi}_i | \Pi_i \otimes I | \tilde{\psi}_i \rangle} ~.
 \label{E:opt_F_cond_pre}
\end{eqnarray}
 Note that the function $g(x) = (a + b x)^2/(a^2 + \nu x)$ is convex for any
 $a,b,\nu,x\geq 0$. So by applying Jensen's inequality to the right hand side
 of Eq.~(\ref{E:opt_F_cond_pre}) and by using Eq.~(\ref{E:Def_p_max}), I
 conclude that
\begin{eqnarray}
 \min_\text{Alice} \max_\text{Bob} \bar{F}_\text{cond}(p,p_\text{max}) & \geq &
  \frac{\left( a + b p_\text{max} \right)^2}{a^2 + \nu p_\text{max}} \nonumber
  \\
 & = & \frac{\left( a + b p_\text{max} \right)^2}{p} ~. \label{E:opt_F_cond}
\end{eqnarray}

\subsection{The optimality of the measurement strategy with respected to the
 average fidelity performance measure}
\label{Subsec:optimal}
 For an arbitrary quantum seal chosen by Alice, it may be possible to find a
 POVM measurement ${\mathcal E}$, whose probability of correctly determining
 the sealed message equals $p$, satisfying $\bar{F}({\mathcal E}) >
 \bar{F}({\mathcal E}_{p,p_\text{max}})$. However, Alice may choose the quantum
 seal reported in the next paragraph. It turns out that the value of
 $\bar{F}({\mathcal E})$ for this seal is upper-bounded by the right hand side
 of Eq.~(\ref{E:opt_F}). This makes ${\mathcal E}_{p,p_\text{max}}$ an optimal
 measurement strategy for Bob when using $\bar{F}$ as the performance
 indicator.

 Consider the quantum sealing scheme with $\eta_i = 1/N$ and
\begin{eqnarray}
 |\tilde{\psi}_i\rangle & = & p_\text{max}^{1/2} |i\rangle_\text{B} \otimes
  |i\rangle_\text{A} \otimes |i\rangle_\text{A} \nonumber \\
 & & ~+ \sqrt{\frac{1 - p_\text{max}}{N-1}} \sum_{j\neq i} |j\rangle_\text{B}
  \otimes |j\rangle_\text{A} \otimes |i\rangle_\text{A} \label{E:opt_scheme}
\end{eqnarray}
 for $i=0,1,\ldots ,N-1$, where each of the three quantum registers used in the
 above scheme is $N$-dimensional with basis $\{ |j\rangle \}_{j=0}^{N-1}$. It
 is straight-forward to check that $|\tilde{\psi}_i\rangle$'s are orthonormal
 and that
\begin{eqnarray}
 \rho_i & = & \text{Tr}_\text{A} (|\tilde{\psi}_i\rangle\langle\tilde{\psi}_i|)
  = p_\text{max} |i\rangle\langle i| + \frac{1-p_\text{max}}{N-1} (I -
  |i\rangle\langle i| ) \nonumber \\
 & = & \frac{(p_\text{max} N - 1) |i\rangle\langle i| + (1 - p_\text{max})
  I}{N-1} \nonumber \\
 & \equiv & c(p_\text{max},N) |i\rangle\langle i| + d(p_\text{max},N) I
  \nonumber \\
 & \equiv & c |i\rangle\langle i| + d I ~. \label{E:rho_i_opt_scheme}
\end{eqnarray}

 Now I show that this sealing scheme is the most stringent one in the sense
 that the resultant fidelity of the quantum state after any POVM measurement by
 Bob is upper-bounded by the right hand side of Eq.~(\ref{E:opt_F}). Recall
 that ${\mathcal E}$ can be written as $\sum_{i=0}^{N-1} {\mathcal L}_i$ where
 ${\mathcal L}_i$ is the superoperator describing Bob's action when he
 concludes that the sealed message is $i$. In general, the action of each
 ${\mathcal L}_i$ on a density matrix $\rho$ can be written as
\begin{equation}
 {\mathcal L}_i (\rho) = \sum_j Q_{ij} \rho Q_{ij}^\dag ~. \label{E:Def_L_i}
\end{equation}
 Clearly, $Q_{ij}$'s satisfy
\begin{equation}
 \sum_{i,j} Q_{ij}^\dag Q_{ij} = I \label{E:Q_ij_constraint1}
\end{equation}
 and
\begin{equation}
 \frac{1}{N} \sum_{i,j} \text{Tr} ( Q_{ij}^\dag Q_{ij} \rho_i ) = p ~.
 \label{E:Q_ij_constraint2}
\end{equation}
 From Eqs.~(\ref{E:Def_F}) and~(\ref{E:opt_scheme}), the average fidelity of
 the state after Bob has applied the POVM ${\mathcal E}$ is given by
\begin{eqnarray}
 \bar{F} & = & \frac{1}{N} \sum_{i,j,k} \left| c \langle i| Q_{jk} | i\rangle +
  d \text{Tr} ( Q_{jk} ) \right|^2 \nonumber \\
 & = & \frac{c^2}{N} \sum_{i,j,k} \left| \langle i| Q_{jk} |i\rangle \right|^2
  + d(d+\frac{2c}{N}) \sum_{j,k} \left| \text{Tr} (Q_{jk}) \right|^2 \nonumber
  \\
 & \leq & \frac{c^2}{N} \sum_{i,j,k,m} \langle i | Q_{jk}^\dag |m\rangle\langle
  m| Q_{jk} |i\rangle \nonumber \\
 & & ~~+ d(d+\frac{2c}{N}) \sum_{j,k} \left| \text{Tr} (Q_{jk}) \right|^2
  \nonumber \\
 & = & c^2 + d(d+\frac{2c}{N}) \sum_{j,k} \left| \text{Tr} (Q_{jk}) \right|^2
  ~, \label{E:F_bar_E_resultant}
\end{eqnarray}
 where the equality holds if and only if $\langle i | Q_{jk} |m\rangle = 0$
 for all $i\neq m$. Subjected to the constraints in
 Eqs.~(\ref{E:Q_ij_constraint1}) and~(\ref{E:Q_ij_constraint2}), the last line
 of Eq.~(\ref{E:F_bar_E_resultant}) is maximized if
\begin{eqnarray}
 Q_{i0} & = & \sqrt{\frac{p_\text{max} - p + N p - 1}{p_\text{max} N - 1}}
  |i\rangle\langle i| \nonumber \\
 & & ~+ \sqrt{\frac{p_\text{max} - p}{p_\text{max} N - 1}} \sum_{j\neq i}
  |j\rangle\langle j| \nonumber \\
 & = & a(\nu,N) I + b(\nu,N) |i\rangle\langle i| ~, \label{E:Opt_M_i}
\end{eqnarray}
 and $Q_{ij} = 0$ for all $j\neq 0$. (Note that although this set of $Q_{ij}$'s
 is not unique, it is straight-forward to check that all $Q_{ij}$'s that
 maximizes $\bar{F}$ are equivalent in the sense that they give the same POVM
 ${\mathcal E}$.) In addition, the maximum probability $p_\text{max}$ of
 distinguishing $\rho_i$'s is attained by the measurement operators $|i\rangle
 \langle i|$'s as these operators satisfy Eqs.~(\ref{E:opt_dist1})
 and~(\ref{E:opt_dist2}). Therefore, $Q_{i0} = M_i$ for all $i$ and hence
 ${\mathcal E} = {\mathcal E}_{p,p_\text{max}}$. Consequently, for this
 particular quantum seal chosen by Alice, $\bar{F}$ is at most equal to the
 right hand side of Eq.~(\ref{E:opt_F}). Besides, such an equality can be
 obtained by using the POVM ${\mathcal E}_{p,p_\text{max}}$. That is to say,
 ${\mathcal E}_{p,p_\text{max}}$ is an optimal measurement strategy for Bob
 with respected to the average fidelity performance measure when Alice uses the
 sealing scheme in Eq.~(\ref{E:opt_scheme}).

\subsection{Analysis of the measurement strategy with respected to the average
 conditional fidelity performance measure}
\label{Subsec:Analysis_opt_bar_F_cond}

 Using the same notation as in Subsec.~\ref{Subsec:optimal}, the average
 conditional fidelity of the state after Bob has applied the POVM
 ${\mathcal E}$ to the quantum seal in Eq.~(\ref{E:opt_scheme}) equals
\begin{equation}
 \bar{F}_\text{cond} = \frac{1}{N} \sum_{i=0}^{N-1} \frac{\sum_j \left| c
 \langle i | Q_{ij} | i\rangle + d \text{Tr} (Q_{ij}) \right|^2}{\sum_j \left[
 c \langle i | Q_{ij}^\dag Q_{ij} |i\rangle + d \text{Tr}(Q_{ij}^\dag Q_{ij})
 \right]} ~. \label{E:F_bar_cond_E_resultant}
\end{equation}

 By constrained extremization, it is easy to show that for a fixed value of
 $\sum_j  \left[ c \langle i | Q_{ij}^\dag Q_{ij} | i\rangle + d \text{Tr} (
 Q_{ij}^\dag Q_{ij} ) \right]$, the $i$th term in the above equation is
 maximized if (1)~$Q_{ij} = 0$ for all $j\neq 0$, (2)~$|k\rangle$ is an
 eigenvector of $Q_{i0}$ whose eigenvalue $\tau_{ik} \geq 0$ for all $k$,
 (3)~$\tau_{ik} = \tau_{ik'}$ for all $k,k' \neq i$, and (4)~$\tau_{ii} \geq
 \tau_{ik}$ for all $k\neq i$. However, one cannot jump to the conclusion that
 $\bar{F}_\text{cond}$ is maximized by picking $Q_{i0} = a(\nu,N) I + b(\nu,N)
 |i\rangle\langle i|$. Actually, this conclusion is wrong in general. A
 counterexample is given below: let $N = 8$, $p_\text{max} = 0.9$ and $p =
 0.3$. By choosing $Q_{i0} = a(\nu,N) I + b(\nu,N) |i\rangle\langle i|$,
 $\bar{F}_\text{cond} = 0.980204$. In contrast, by choosing $Q_{i0} = 3 ( 2
 \sqrt{47} |i\rangle\langle i| + \sqrt{13} \sum_{j\neq i} |j\rangle\langle j|)
 / 15\sqrt{31}$ for $i=0,1,\ldots ,5$ and $Q_{i0} = ( 2\sqrt{109} |i\rangle
 \langle i| + 3\sqrt{29} \sum_{j\neq i} |j\rangle\langle j| ) / 5\sqrt{31}$ for
 $i=6,7$, then $\bar{F}_\text{cond} = 0.981247$. That is to say, for
 $p_\text{max} = 0.9$ and $p = 0.3$, the $\bar{F}_\text{cond}$ caused by a
 certain asymmetric set of Kraus operators $\{ Q_{i0} \}$ (in the sense that
 there exist $i,j$ such that $Q_{i0} \neq U Q_{j0} U^{-1}$ for some permutation
 operation $U$ of the standard basis) is greater than that caused by a
 symmetric set of Kraus operators. In fact, this symmetry breaking phenomenon
 is partly due to the fact that $\bar{F}_\text{cond}({\mathcal E})$, unlike
 $\bar{F}({\mathcal E})$, is not a linear function of ${\mathcal E}$. And the
 nonlinear dependence of $\bar{F}_\text{cond}$ on ${\mathcal E}$ makes the
 determination of $\min_\text{Alice} \max_\text{Bob} \bar{F}_\text{cond}
 (p,p_\text{max})$ difficult.

 In spite of this difficulty, the function $\min_\text{Alice} \max_\text{Bob}
 \bar{F}_\text{cond}(p,p_\text{max})$ can be found in the following three
 special cases: (1)~$N\leq 5$, (2)~$p = 1/N$ and (3)~$p = p_\text{max}$.

 For the first case ($N\leq 5$), by constrained maximization, the $i$th term in
 Eq.~(\ref{E:F_bar_cond_E_resultant}) is upper-bounded by $h(x_i) = [a(\nu(x_i,
 p_\text{max},N),N) + b(\nu(x_i,p_\text{max},N),N) p_\text{max}]^2 / x_i$,
 where $x_i = \sum_j \left[ c \langle i | Q_{ij}^\dag Q_{ij} | i\rangle + d
 \text{Tr} (Q_{ij}^\dag Q_{ij}) \right] / \sum_j \text{Tr} (Q_{ij}^\dag
 Q_{ij})$. Observe that $h(x)$ is concave for $N\leq 5$ and $p\in [1/N,
 p_\text{max}]$. (One way to see this is to use Mathematica to check that
 $h''(p) \leq 0$.) Therefore, Eq.~(\ref{E:F_bar_cond_E_resultant}) is upper
 bounded by $h(p) = (a + b p_\text{max})^2 / p$ if $N\leq 5$. Surely, this
 upper bound is attained by picking the POVM ${\mathcal E}_{p,p_\text{max}}$,
 namely, the one that also maximizes the performance indicator $\bar{F}$.

 The second case ($p = 1/N$) is trivial as Eq.~(\ref{E:opt_F_cond}) implies
 that the average conditional fidelity of the state after applying
 ${\mathcal E}_{1/N,p_\text{max}}$ equals $1$.

 For the third case ($p = p_\text{max}$), the symmetry of the quantum seal in
 Eq.~(\ref{E:opt_scheme}) demands that the denominator in each term of the sum
 in Eq.~(\ref{E:F_bar_cond_E_resultant}) must all equal to $p_\text{max}$.
 Using the same constrained maximization analysis in the first case, each term
 in Eq.~(\ref{E:F_bar_cond_E_resultant}) is upper-bounded by $h(p_\text{max}) =
 p_\text{max}$. Hence, $\bar{F}_\text{cond} \leq p_\text{max}$ in this case.
 Moreover, this upper bound is attained by the POVM
 ${\mathcal E}_{p_\text{max},p_\text{max}}$.

 In summary, I have proven

\begin{widetext}
 \begin{Thrm}
  Let $1/N \leq p \leq p_\text{max}$. Then,
  \begin{equation}
   \min_\text{Alice} \max_\text{Bob} \bar{F}(p,p_\text{max}) = 1 - \left(
   \sqrt{\nu+\frac{1-\nu}{N}} - \sqrt{\frac{1-\nu}{N}} \right)^2 \left[ 1 -
   p_\text{max}^2 - \frac{(1-p_\text{max})^2}{N-1} \right] \label{E:opt_F_all}
  \end{equation}
  and
  \begin{equation}
   \min_\text{Alice} \max_\text{Bob} \bar{F}_\text{cond} (p,p_\text{max}) \geq
   \frac{\left[ \sqrt{\frac{1-\nu}{N}} + \left( \sqrt{\nu + \frac{1-\nu}{N}} -
   \sqrt{\frac{1-\nu}{N}} \right) p_\text{max} \right]^2}{p} ~,
   \label{E:opt_F_cond_all}
  \end{equation}
  where $\nu$ is given by Eq.~(\ref{E:Def_nu}). In particular,
  Eq.~(\ref{E:opt_F_cond_all}) is an equality if $p = 1/N, p_\text{max}$ or $N
  \leq 5$. Furthermore,
  \begin{equation}
   \min_\text{Alice} \max_\text{Bob} \bar{F} (p_\text{max},p_\text{max}) =
   p_\text{max}^2 + \frac{(1-p_\text{max}^2)}{N-1}
   \label{E:opt_F_all_special}
  \end{equation}
  and
  \begin{equation}
   \min_\text{Alice} \max_\text{Bob} \bar{F}_\text{cond}
   (p_\text{max},p_\text{max}) = p_\text{max} ~.
   \label{E:opt_F_cond_all_special}
  \end{equation}
  \label{Thrm:main}
 \end{Thrm}
\end{widetext}

\begin{figure}[t]
 \centering
 \includegraphics*[scale=0.7]{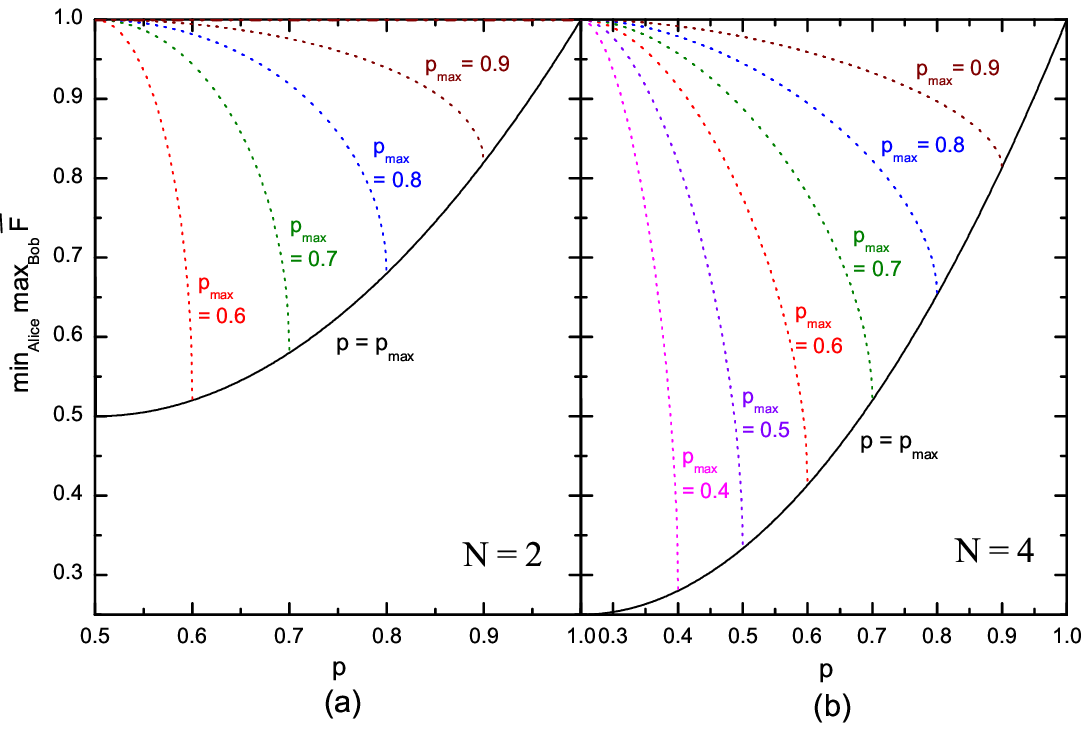}
 \caption{(Color online) Dotted curves show $\min_\text{Alice} \max_\text{Bob}
  \bar{F}(p,p_\text{max})$ \emph{vs.} $p$ for various values of $p_\text{max}$
  with (a)~$N=2$ and (b)~$N=4$. The solid curves show the case of $p =
  p_\text{max}$.}
 \label{F:fidelity}
\end{figure}

\begin{figure}[t]
 \centering
 \includegraphics*[scale=0.7]{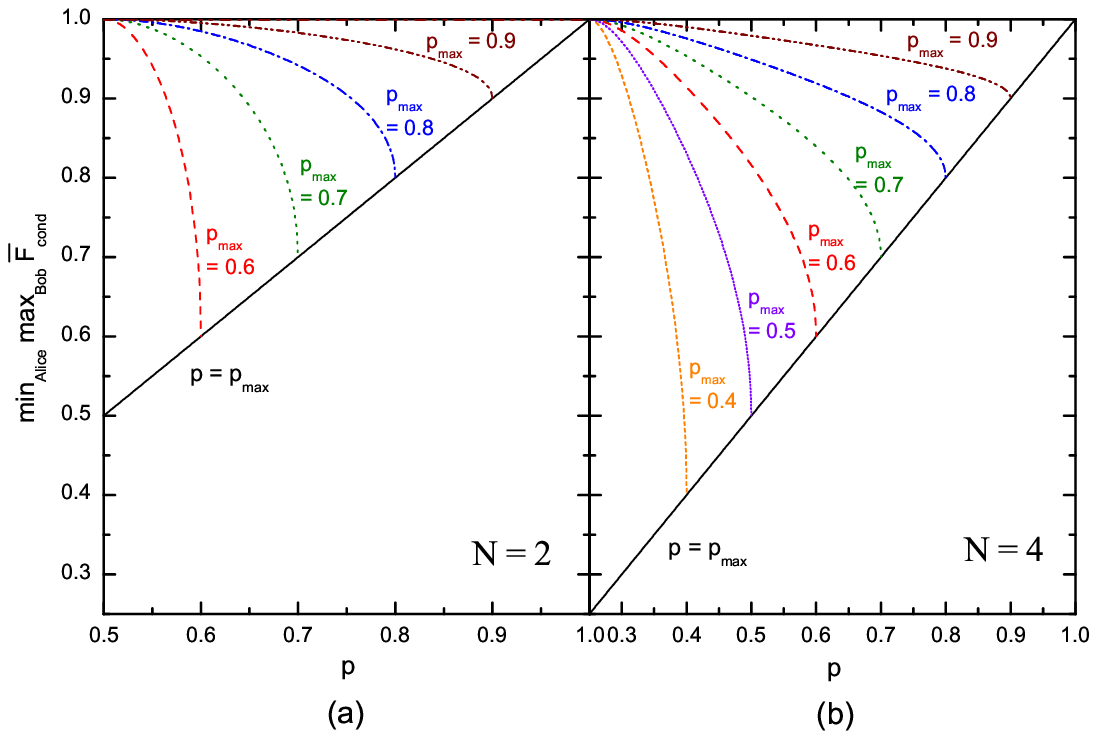}
 \caption{(Color online) Dotted curves show $\min_\text{Alice} \max_\text{Bob}
  \bar{F}_\text{cond} (p,p_\text{max})$ \emph{vs.} $p$ for various values of
  $p_\text{max}$ with (a)~$N=2$ and (b)~$N=4$. The solid curves show the case
  of $p = p_\text{max}$.}
 \label{F:fidelity_cond}
\end{figure}

 Figs.~\ref{F:fidelity} and~\ref{F:fidelity_cond} show $\min_\text{Alice}
 \max_\text{Bob} \bar{F} (p,p_\text{max})$ and $\min_\text{Alice}
 \max_\text{Bob} \bar{F}_\text{cond} (p,p_\text{max})$ \emph{vs.} $p$ for
 different $p_\text{max}$ when $N = 2,4$. Note that $\min_\text{Alice}
 \max_\text{Bob} \bar{F} (p,p_\text{max})$ and $\min_\text{Alice}
 \max_\text{Bob} \bar{F}_\text{cond} (p,p_\text{max})$ are discontinuous at $p
 = 1/N$ whenever $p_\text{max} > 1/N$. This discontinuity originates from the
 sudden change in the dimension of the set $\{ \rho_i \}_{i=0}^{N-1}$ around
 the point $p = 1/N$ \cite{bitsealimposs}. Besides, $\min_\text{Alice}
 \max_\text{Bob} \bar{F} (p,p_\text{max}) \rightarrow 1 - p (1-p_\text{max}^2)
 / p_\text{max}$. Note also that $\min_\text{Alice} \max_\text{Bob} \bar{F} (p,
 p_\text{max})$ is a concave function of $p$ for a fixed value of
 $p_\text{max}$ for $\bar{F} ({\mathcal E})$ is a linear function of
 ${\mathcal E}$.

 Finally, I remark that derivations of the upper and lower bounds for
 $\min_\text{Alice} \max_\text{Bob} \bar{F} (p,p_\text{max})$ and
 $\min_\text{Alice} \max_\text{Bob} \bar{F}_\text{cond} (p,p_\text{max})$
 reported in this Section are also valid in the case of determining partial
 information on the original message via the partition ${\mathfrak P}$.

\section{Proof of insecurity of quantum seal}
\label{Sec:Insecurity}

 Although the functional form of $\min_\text{Alice} \max_\text{Bob}
 \bar{F}_\text{cond}(p,p_\text{max})$ is not known for $N>5$ and $1/N < p <
 p_\text{max}$, its lower bound stated in Theorem~\ref{Thrm:main} is already
 sufficiently stringent to help proving the insecurity of quantum seal.
 Specifically, to fix $\nu = 1/2$ by choosing the appropriate $p$,
 Theorem~\ref{Thrm:main} implies the existence of a POVM measurement
 ${\mathcal E}_{p,p_\text{max}}$ that make both $\min_\text{Alice}
 \max_\text{Bob} \bar{F}$ and $\min_\text{Alice} \max_\text{Bob}
 \bar{F}_\text{cond}$ greater than $1/2$ for all $N\geq 2$. In other words,
 this measurement obtains non-trivial information on the sealed message and
 escapes verifier's detection at least half of the time. Hence, all quantum
 seals are insecure.

 In fact, the major loophole in He's proof of the security of his quantum
 string seal in Ref.~\cite{stringseal} is that he incorrectly assumed that
 measuring all the qubits is the only method to obtain a significant portion of
 information of the sealed message.

 Recently, He proposed the following method of attack \cite{He_attack}: Bob
 measures the sealed message using $\Pi_i$'s as the POVM elements with
 probability $1/2$; and he randomly guesses the sealed message without actually
 measuring otherwise. His mixed strategy escapes verifier's detection at least
 half of the time (as measured by $\bar{F}$) and obtains non-trivial
 information on the sealed message. Nevertheless, $\bar{F}_\text{cond}$ of this
 strategy approaches $p_\text{max}/2$ as $N\rightarrow\infty$. Compared with
 Theorem~\ref{Thrm:main}, the average conditional fidelity of He's attack is
 only about $1/2$ that of the optimal strategy.

\section{Conclusions}
\label{Sec:Conclusion}
 To summarize, I have extended the study of information disturbance tradeoff
 for quantum bit seal \cite{bitsealimposs} to the case of quantum string seal.
 Specifically, I show that the average fidelity and average conditional
 fidelity of the measured state is greater than or equal to the right hand side
 of Eqs.~(\ref{E:opt_F_all}) and~(\ref{E:opt_F_cond_all}), respectively.
 Furthermore, the equalities are simultaneously attained by a specific quantum
 sealing scheme provided that $N\leq 5$ or $p = 1/N, p_\text{max}$. A
 consequence of this information disturbance tradeoff expression is that all
 quantum seals are insecure provided that one has access to a quantum computer.

 One of the major reasons I can extend the earlier result on quantum bit seal
 in Ref.~\cite{bitsealimposs} here is that I replace the classical $L_1$
 distance by the probability of distinguishing two classical probability
 distributions. The later concept readily extends to the case of $N>2$.
 Actually, it can be shown that for $N=2$ the measurement
 ${\mathcal E}_{p,p_\text{max}}$ together with the expression for
 $\min_\text{Alice} \max_\text{Bob} \bar{F}(p,p_\text{max})$ are the same as
 the ones reported in Ref.~\cite{bitsealimposs}. I also remark that even though
 I consider only the case of sealing finite number of messages, the arguments
 used in this Paper can be easily extended to cover the case of sealing
 infinite number of messages using states in an infinite dimensional Hilbert
 space.

 Although quantum seal is not unconditionally secure, the construction of
 ${\mathcal E}_{p,p_\text{max}}$ requires Bob to find the POVM measurement $\{
 \Pi_i \}$ that distinguish the density matrices $\rho_i$'s with
 \emph{a priori} probability $\eta_i$'s with minimum error. In general, it is
 difficult to explicitly find the $\Pi_i$'s; and a quantum computer is needed
 to implement ${\mathcal E}_{p,p_\text{max}}$. So, it may be possible to
 construct a quantum seal that is secure under certain computational or
 hardware assumptions. Last but not least, it is instructive to find
 $\min_\text{Alice} \max_\text{Bob} \bar{F}_\text{cond}$ for $N > 5$ and $1/N <
 p < p_\text{max}$.

\begin{acknowledgments}
 Useful discussions with K.~H. Ho is gratefully acknowledged. This work is
 supported by the RGC grant HKU~7010/04P of the HKSAR Government.
\end{acknowledgments}

\bibliography{qc36.3}

\begin{thebibliography}{13}
\expandafter\ifx\csname natexlab\endcsname\relax\def\natexlab#1{#1}\fi
\expandafter\ifx\csname bibnamefont\endcsname\relax
  \def\bibnamefont#1{#1}\fi
\expandafter\ifx\csname bibfnamefont\endcsname\relax
  \def\bibfnamefont#1{#1}\fi
\expandafter\ifx\csname citenamefont\endcsname\relax
  \def\citenamefont#1{#1}\fi
\expandafter\ifx\csname url\endcsname\relax
  \def\url#1{\texttt{#1}}\fi
\expandafter\ifx\csname urlprefix\endcsname\relax\def\urlprefix{URL }\fi
\providecommand{\bibinfo}[2]{#2}
\providecommand{\eprint}[2][]{\url{#2}}

\bibitem[{\citenamefont{Bechmann-Pasquinucci}(2003)}]{quantseal}
\bibinfo{author}{\bibfnamefont{H.}~\bibnamefont{Bechmann-Pasquinucci}},
  \bibinfo{journal}{Int. J. Quant. Inform.} \textbf{\bibinfo{volume}{1}},
  \bibinfo{pages}{217} (\bibinfo{year}{2003}).

\bibitem[{\citenamefont{Chau}()}]{oldseal}
\bibinfo{author}{\bibfnamefont{H.~F.} \bibnamefont{Chau}},
  \emph{\bibinfo{title}{Sealing quantum message by quantum code}},
  \bibinfo{note}{quant-ph/0308146}.

\bibitem[{\citenamefont{Singh and Srikanth}(2005)}]{anotherseal}
\bibinfo{author}{\bibfnamefont{S.~K.} \bibnamefont{Singh}} \bibnamefont{and}
  \bibinfo{author}{\bibfnamefont{R.}~\bibnamefont{Srikanth}},
  \bibinfo{journal}{Physica Scripta} \textbf{\bibinfo{volume}{71}},
  \bibinfo{pages}{433} (\bibinfo{year}{2005}).

\bibitem[{\citenamefont{He}({\natexlab{a}})}]{anotherseal2}
\bibinfo{author}{\bibfnamefont{G.-P.} \bibnamefont{He}},
  \emph{\bibinfo{title}{Quantum secret sharing, hiding and sealing of classical
  data against collective measurement}},
  \bibinfo{howpublished}{quant-ph/0502091v1}.

\bibitem[{\citenamefont{He}(2006)}]{stringseal}
\bibinfo{author}{\bibfnamefont{G.-P.} \bibnamefont{He}}, \bibinfo{journal}{Int.
  J. Quant. Inform.} \textbf{\bibinfo{volume}{4}}, \bibinfo{pages}{677}
  (\bibinfo{year}{2006}).

\bibitem[{\citenamefont{Bechmann-Pasquinucci
  et~al.}(2005)\citenamefont{Bechmann-Pasquinucci, D'Ariano, and
  Macchiavello}}]{impossibleperfect}
\bibinfo{author}{\bibfnamefont{H.}~\bibnamefont{Bechmann-Pasquinucci}},
  \bibinfo{author}{\bibfnamefont{G.~M.} \bibnamefont{D'Ariano}},
  \bibnamefont{and}
  \bibinfo{author}{\bibfnamefont{C.}~\bibnamefont{Macchiavello}},
  \bibinfo{journal}{Int. J. Quant. Inform.} \textbf{\bibinfo{volume}{3}},
  \bibinfo{pages}{435} (\bibinfo{year}{2005}).

\bibitem[{\citenamefont{He}(2005)}]{bound}
\bibinfo{author}{\bibfnamefont{G.-P.} \bibnamefont{He}},
  \bibinfo{journal}{Phys. Rev. A} \textbf{\bibinfo{volume}{71}},
  \bibinfo{pages}{054304} (\bibinfo{year}{2005}).

\bibitem[{\citenamefont{Chau}(2006)}]{bitsealimposs}
\bibinfo{author}{\bibfnamefont{H.~F.} \bibnamefont{Chau}},
  \bibinfo{journal}{Phys. Lett. A} \textbf{\bibinfo{volume}{354}},
  \bibinfo{pages}{31} (\bibinfo{year}{2006}).

\bibitem[{\citenamefont{Buhrman et~al.}({\natexlab{a}})\citenamefont{Buhrman,
  Christandl, Hayden, Lo, and Wehner}}]{bitstringcommitment}
\bibinfo{author}{\bibfnamefont{H.}~\bibnamefont{Buhrman}},
  \bibinfo{author}{\bibfnamefont{M.}~\bibnamefont{Christandl}},
  \bibinfo{author}{\bibfnamefont{P.}~\bibnamefont{Hayden}},
  \bibinfo{author}{\bibfnamefont{H.-K.} \bibnamefont{Lo}}, \bibnamefont{and}
  \bibinfo{author}{\bibfnamefont{S.}~\bibnamefont{Wehner}},
  \emph{\bibinfo{title}{On the (im)possibility of quantum string commitment}},
  \bibinfo{note}{quant-ph/0504078}.

\bibitem[{\citenamefont{Buhrman et~al.}({\natexlab{b}})\citenamefont{Buhrman,
  Christandl, Hayden, Lo, and Wehner}}]{bitstringcommitment2}
\bibinfo{author}{\bibfnamefont{H.}~\bibnamefont{Buhrman}},
  \bibinfo{author}{\bibfnamefont{M.}~\bibnamefont{Christandl}},
  \bibinfo{author}{\bibfnamefont{P.}~\bibnamefont{Hayden}},
  \bibinfo{author}{\bibfnamefont{H.-K.} \bibnamefont{Lo}}, \bibnamefont{and}
  \bibinfo{author}{\bibfnamefont{S.}~\bibnamefont{Wehner}},
  \emph{\bibinfo{title}{Security of quantum bit string commitment depends on
  the information measure}}, \bibinfo{note}{quant-ph/0609237, to appear in
  Phys. Rev. Lett.}

\bibitem[{\citenamefont{Holevo}(1973)}]{opt1}
\bibinfo{author}{\bibfnamefont{A.~S.} \bibnamefont{Holevo}},
  \bibinfo{journal}{J. Multivar. Anal.} \textbf{\bibinfo{volume}{3}},
  \bibinfo{pages}{337} (\bibinfo{year}{1973}).

\bibitem[{\citenamefont{Yuen et~al.}(1975)\citenamefont{Yuen, Kennedy, and
  Lax}}]{opt2}
\bibinfo{author}{\bibfnamefont{H.~P.} \bibnamefont{Yuen}},
  \bibinfo{author}{\bibfnamefont{R.~S.} \bibnamefont{Kennedy}},
  \bibnamefont{and} \bibinfo{author}{\bibfnamefont{M.}~\bibnamefont{Lax}},
  \bibinfo{journal}{IEEE Trans. Inform. Theo.} \textbf{\bibinfo{volume}{21}},
  \bibinfo{pages}{125} (\bibinfo{year}{1975}).

\bibitem[{\citenamefont{He}({\natexlab{b}})}]{He_attack}
\bibinfo{author}{\bibfnamefont{G.-P.} \bibnamefont{He}},
  \emph{\bibinfo{title}{Secure quantum string seal exists}},
  \bibinfo{howpublished}{quant-ph/0602159}.

\end{thebibliography}
\end{document}